\documentclass{aa}  
\usepackage{graphicx}
\usepackage{color}
\usepackage[varg]{txfonts}
\begin{document}

   \title{Drifting of the line-tied footpoints of CME flux-ropes}


   \author{Guillaume Aulanier\inst{1}
          \and
          Jaroslav Dud\'{\i}k\inst{2}
          }

   \institute{LESIA, Observatoire de Paris, Universit\'e PSL, CNRS, Sorbonne Universit\'e, 
	            Universit\'e Paris Diderot, 5 place Jules Janssen, 92190 Meudon, France\\
              \email{guillaume.aulanier@obspm.fr}
         \and Astronomical Institute of the Czech Academy of Sciences, Fri\v{c}ova 298, 
				      251 65 Ond\v{r}ejov, Czech Republic\\
             \email{jaroslav.dudik@asu.cas.cz}
             }

   \date{Received 11 sept 2018 ; accepted 8 nov 2018}

 
  \abstract
  {Bridging the gap between heliospheric and solar observations of eruptions requires to map ICME footpoints down to the Sun's surface. But this not straightforward. Improving the understanding of the spatio-temporal evolutions of eruptive flares requires a comprehensive standard model. But the current one is two-dimensional only and it cannot address the question of CME footpoints. 
	}
   {Existing 3D extensions to the standard model show that flux-rope footpoints are surrounded by curved-shaped QSL-footprints that can be related with hook-shaped flare-ribbons. We build upon this finding and further address the joint questions of their time-evolution, and of the formation of flare loops at the ends of flaring PILs of the erupting bipole, which are both relevant for flare understanding in general and for ICME studies in particular. 
	 }
   {We calculate QSLs and relevant field lines in an MHD simulation of a torus-unstable flux-rope. The evolving QSL footprints are used to define the outer edge of the flux rope at different times, and to identify and characterize new 3D reconnection geometries and sequences that occur above the ends of the flaring PIL. We also analyse flare-ribbons as observed in EUV by {\it SDO}/AIA and {\it IRIS} during two X-class flares.  
	 }
   {The flux-rope footpoints are drifting during the eruption, which is unexpected due to line-tying. This drifting is due to a series of coronal reconnections that erode the flux rope on one side and enlarge it on the other side. Other changes in the flux-rope footpoint-area are due to multiple reconnections of individual field lines whose topology can evolve sequentially from arcade to flux rope and finally to flare loop. These are associated with deformations and displacements of QSL footprints, which resemble those of the studied flare ribbons. 
	 }
   {Our model predicts continuous deformations and a drifting of ICME flux-rope footpoints whose areas are  surrounded by equally-evolving hooked-shaped flare-ribbons, as well as the formation of flare loops at the ends of flaring PILs which originate from the flux-rope itself, both of which being due to purely three-dimensional reconnection geometries. The observed evolution of flare-ribbons in two events supports the model, but more observations are required to test all its predictions. 
	 }


\authorrunning{G. Aulanier and 
               J. Dud\'{\i}k
					    }

   \maketitle
%

\section{Introduction}
\label{secintro}

Solar eruptions are the most energetic manifestations of the Sun's magnetic activity \citep{Schrij12,BS17}. All together their flares and coronal mass ejections (CMEs) are among the most important drivers of space weather \citep{Schwenn06,Boc18}. Characterizing the magnetic linkages of interplanetary ejections (ICMEs) from the heliosphere, in which they propagate, down to the lower layers of the solar atmosphere, in which they originate, is one of the key steps towards their global understanding. However these linkages are still difficult to establish. One major reason is that observations of the solar atmosphere rely on remote-sensing instruments only, whereas interplanetary observations mostly rely on in-situ measurements. Thanks to the increasing capacities of the instruments, to the recent availability of heliospheric imagers, and to the development of global numerical models, associations have been made between solar eruptions, ICMEs, and their respective large-scale properties \citep{Howard07,Demou08,Forest13,Man17}. Nevertheless, fully bridging the gap such as e.g. mapping an interplanetary CME flux-tube crossed by a spacecraft like the upcoming {\em Solar Orbiter} down to its footpoint at the Sun's surface or to its pre-eruptive state in the corona remains a challenge. 

The solar feature that is most commonly associated with ICME footpoints is transient coronal holes (TCHs) also called coronal dimmings \citep{Ster97,Thom98,WebbJ00}. When an eruption occurs, TCHs develop at various distances from the flaring polarity-inversion line (PIL). Their intensity decreases in EUV and SXR are attributed to a coronal density decrease due to the loop expansions during the CME. Unfortunately their direct linkage to ICME flux-ropes has been shown not to be straightfoward. One reason is that TCHs do not only develop around the flaring PIL, but also in nearby active regions \citep{Kahler01} and the magnetic flux summed over the whole dimmings tends to be incompatible with the axial flux of the corresponding ICME flux-rope \citep{Mandri07}. These may be explained by the fact that TCH areas and fluxes are largely dominated by their components located far away from the flaring PIL \citep{Diss18}, and that these so-called secondary dimmings may form at the footpoints of expanding loops that are carried along with the CME, either because they reconnect with the erupting flux-rope \citep{Gibss08, Cohen10}, or because they are merely pushed from below by the eruption \citep{DelHA07}. The so-called core (or twin) dimmings that develop close to the eruption site may thus be considered as the footpoints of the ICME (or the pre-eruptive) flux rope. But even this is not obvious because twin dimmings are simply not always observed \citep{Diss18}, and because their dynamical nature questions their link with interplanetary flux-ropes \citep[as raised by][]{Kahler01}. Indeed twin dimmings can be very dynamic. Firstly their outer boundaries spread out to distances that can be larger than the erupting active region \citep{Thom98,LiuLee07}. And secondly their ``boundaries closer to the magnetic neutral line generally move away from it as the closed-loop X-ray arcades expand [...] the TCHs tend to disappear only by a net contraction of the boundaries'' \citep[to quote][]{Kahler01}. Therefore coronal dimmings can still not be readily used to locate the footpoints of the ICME and the pre-eruptive flux-rope. 

Recent observations show that the time-evolution of twin dimmings is associated with flare ribbons \citep[as can be seen in][]{LiuLee07,Warren11,Cheng16} and in particular at the ends of the flaring PILs, where the ribbons curve into hook shapes. The latter have been reported for many events \citep[e.g. in][]{Chandra09,Aula12,Schrijver11,Dudik14,Dudik16}. Linking these hooked ribbons with flux ropes is impossible in the frame of the standard flare (or CSHKP) model \citep{Car64,Stu66,Hir74,Kop76}. The reason is that the model is two-dimensional and invariant by translation. So it describes flare ribbons in which reconnected loops are rooted as two parallel lines, which have no ending. And it describes the erupting flux-rope as a detached plasmoid, which has no footpoint. Thus one should consider tree-dimensional extensions of the CSHKP flare model in which the erupting flux rope is anchored at both ends \citep[as reviewed in][]{Janvier15}. In this geometry the flux rope is surrounded by sharp gradients of field-line connectivities known as quasi-separatrix layers \citep[QSLs, as defined by][]{Priest95,Dem96a}. The QSL footprints have a double-J shape whose curved endings mark out the outer edges of the flux-rope footpoints \citep[as first found by][]{Dem96b}. This property is robust for various static \citep{Titov07,Titov08,Sav12,Par12,Zhao16} and dynamic \citep{Jan13,Jan16,Inoue15,Sav15,Sav16} flux-rope models. When the latter were compared with observations, the curved QSL footprints were associated with the curved ends of hooks of the eruptive-flare ribbons. 
So, as long as an eruptive flux-rope has not reconnected with remote loops rooted in distant active-regions or coronal holes, one could argue that the line-tying should conserve the ICME flux-rope footpoints at their original locations, which are surrounded by the hooks of flare ribbons at onset time of the eruption. 
But this would ignore the possible effects of long-duration couplings between the flare dynamics and the footpoints of the CME, as implied by reports of flare loops gradually expanding into coronal dimming areas \citep{Kahler01,Cheng16}, and of several correlations between flare ribbons and dimmings \citep{Diss18}. Such couplings which can happen at the ends of flaring PILs, however, have not been addressed yet by the 3D extensions of the standard flare model. 

Understanding the physical processes that at work at the endings of flaring PILs in general, and developing refined proxys to follow the locations of ICME footpoints at the Sun's surface in particular, all motivate further developments to the 3D standard model. This is the object of this paper, which reports on new analyzes of an existing MHD simulation for eruptive flares \citep{Zucca15}. 

In Sect.~\ref{sec2}, we briefly describe the numerical simulation, and we explain the methods used to characterize the edge of the flux rope and to identify field lines involved in pairwise reconnections. 
In Sect.~\ref{secrope}, we describe the deformation and drifting of the erupting flux-rope footpoints, and we relate them with reconnection-driven QSL-footprint evolution and flare-loop formation. 
In Sect.~\ref{secreco}, we analyse different types of pairwise reconnections that occur during the eruption, we identify new reconnection geometries and sequences that do not exist in the CSHKP model, and we associate them with the drifting of the flux rope and with the formation of flare loops at the ends of flaring PILs.  
In Sect.~\ref{secobs}, we show the dynamics of flare ribbon hooks in two well-known X-class eruptive flares, and we relate their evolution with that of the modeled QSL footprints. 
Finally, in Sect.~\ref{secccl} we summarize our results in the frame of the development of the 3D standard flare model and of the mapping of ICME flux ropes down to the Sun's surface.

\section{Analyzing an MHD model}
\label{sec2}

\subsection{The line-tied MHD simulation}
\label{secmethmhd}

In this work we use the results of one of the three-dimensional MHD simulations from \citet{Zucca15}. 
The simulations were calculated with the visco-resistive OHM code \citep{Aula05a,Zucca15} with the line-tying and zero-$\beta$ approximations, following a similar methodology as in \citet{Aula10}. 
Initial conditions were built with an asymmetric current-free bipolar magnetic field, and with uniform Alfv\'en speed. Non-dimensionalized units were used: the space-unit was defined as half of the distance $=2$ between two photospheric flux concentrations; the time-unit $ t_\mathrm{A}$ was defined as half the propagation time of an Alfv\'en wave from one polarity center to the other; and the magnetic permeability was set to $\mu=1$. 
Sub-Alfv\'enic narrow shearing motions and large-scale converging motions were kinematically prescribed at the line-tied photospheric boundary at $z=0$. When combined with a finite photospheric resistivity, these motions led to flux-cancellation and quasi-static flux-rope formation at the PIL \citep{vanBalle89}. Owing to asymmetry of the system, one of the flux rope footpoints ended up being rooted near the center of the strongest polarity, while its other footpoint developed at the border of the other polarity in a relatively weak field area. In all simulations the rope eventually became torus-unstable, the photospheric drivers were then stopped, and the eruption generated a line-tied CME expansion. We refer the reader to \citet{Zucca15} for more details. 

In each simulation, a long reconnecting current-sheet formed below the flux rope and produced flare loops through slip-running reconnection. A pair of large-scale coronal vortices developed on the flanks of the flux rope and produced so-called imploding loops \citep{Zucca17,Dudik17}. 
{\color{black} And as in our earlier similar models \citep{Aula12,Jan13,Jan14} reconnection generated current-ribbons at the line-tied boundary that moved away from each other.}
All simulations were eventually halted several tens of Alfv\'en times after the onset of the eruption, due to the sudden and local development of numerical instabilities in the flare current sheet. 

Here we only consider the ``Run D2''. We restrict the analysis from the time of the onset of the eruption at the time $t = 165 t_\mathrm{A}$, up to the end of the simulation which occurs at $t = 244 t_\mathrm{A}$ but at which the flare is still ongoing. We focus our attention on a fraction of the domain that contains the flux rope ($x\in[-4;3.5]$ and $y\in[-6;4]$). We have also checked that our results remain valid overall in the other simulations, in which different pre-eruptive boundary drivings resulted in qualtitatively-similar but quantitatively-different magnetic configurations and eruption dynamics. 

\subsection{Characterizing the outer edges of the flux rope}
\label{secmethqsl}

To study the evolution of the anchorage of the erupting flux rope, several options are possible. Different criteria can be applied to selecting different set of individual field lines and following their time-evolution \citep{Gibss08}. Contrary to that, here we use a method that relies on the identification of the edge of the erupting flux rope. This can be achieved by trial-and-error field-line plotting. But this manual method is not the most accurate, and it does not allow a smooth and quantitative representation of the flux rope boundary. So instead we consider the topological approach as described in Sect.~\ref{secintro} which is based on quasi-separatrix layers (QSLs) across which the field-line connectivity has sharp but finite gradients. 
%

We use the TOPOTR code \citep{Dem96a,Par12} to calculate the squashing degree $Q(z=0)$ whose narrow and high-$Q$ streaks characterize the location of the QSL footprints \citep{Titov02,Aula05b}. Accurate calculations of the squashing degree are numerically demanding, because for higher $Q$ values, which here peak above $10^7$, integration of an increasingly large number field lines around each considered position is required. So we only make the calculation at three selected times of the simulation, i.e. at $t = 164 t_\mathrm{A}$ right before the eruption onset, at $t = 244 t_\mathrm{A}$ just before the end of simulation, and at $t = 204 t_\mathrm{A}$ at midpoint. 

%
%
Double-J shaped QSLs-footprints associated with flux-ropes (see Section~\ref{secintro}) are also known to be associated with narrow concentrations of electric currents \citep{Jan13,Jan14,Jan16} and with observed flare ribbons \citep{Jan14,Jan16,Inoue15,Sav15,Sav16}. These two features, which can be respectively be readily extracted from models or from observations, can thus be used as a proxy for the locations of QSLs. Inspection of the vertical electric-currents at the line-tied boundary $j_z(z=0)$ throughout the simulation did not show any evolution sharper that what can be seen from in the three calculated QSL snapshots. So those are sufficient to capture the overall evolution of the flux-rope footpoint-area. 

At each desired time, the full field-lines that constitute the edge of the flux rope are integrated from a series of footpoints chosen as follows at $z=0$. Firstly, they are placed at the internal edge of the largest of both QSL hooks, which here is located in the relatively weaker negative polarity. And secondly, they are selected at varying space-intervals, chosen to be larger (resp.) smaller in weaker (resp. stronger) $B_z=0$ so that the distances between the field lines are roughly constant at the center of the flux rope. 

\subsection{Identifying pairs of reconnecting field-lines}
\label{secmethreco}

In principle, to understand the changes in connectivity and topology for a field line of particular interest, for example e.g. an arcade field lines that becomes part of the flux rope as in to the standard model, one needs to identify which field line reconnects with it. In other words, reconnecting field line pairs have to be found. 
Unfortunately, this is impossible to achieve rigorously in our simulation that does not contain null points and separatrix surfaces, because of two following properties of three-dimensional finite-B reconnection. 
Firstly, two reconnecting flux tubes do not simply exchange their connections like they do in 2D (and as drawn in several 3D cartoons in the literature). Instead the two tubes break into four, and new connections are created \citep{Priest03}. Secondly, reconnecting field lines do not change their connectivity in a cut-and-paste manner, like they do across separatrices. Instead they continuously reconnect with a series neighboring field lines while they pass through the current sheet, so their footpoints quickly slip along the QSL footprints \citep{Aula06}. 

In spite of these theoretical limitations, field line pairs can be identified in an approximated way. The relatively high values  $10^5<Q<\mathrm{a~few}~10^7$ along most of the length of the QSL footprints in our magnetic field configurations imply the following for a given reconnecting field line \citep{Aula06,Jan13}. Firstly slip-running reconnection occurs during a time-interval much lower than $ t_\mathrm{A}$. During this time field lines footpoints move at super-Alfv\'enic velocities and cover a significant fraction of the total distance between their initial and final state. So field line pairs can almost be identified on short time-scales. However, some additional distance is swept by the field line footpoint during several $t_\mathrm{A}$ preceding and following slipping reconnection regime. Therefore, longer time-scales have to be considered for retrieval of the final positions of field line footpoints. 
In the following we analyze the reconnection of individual field lines between two Alfv\'en times, which is a good compromise so as to cover both regimes while identifying pairs of interacting field lines. 

We follow the same procedure employed (but not described) in \citet{Aula12} to identify pairs of reconnecting field lines $L_a$ and $L_b$. We chose a ``final'' post-reconnection time $t=t_f$. Then we select a footpoint $F_a$ at $z=0$, in the polarity of the broadest of both QSL hooks (here in the negative polarity), so that it is located in the trailing edge of the moving QSL footprint (here we use the current ribbons). Then we integrate the field line $L_a(t_f)$ from $F_a$, and we obtain its conjugate footpoint $F_c$. We also integrate $L_a(t_i)$ from $F_a$ at a given ``initial'' pre-reconnection time $t=t_i<t_f$ (here with $t_f - t_i = 2 t_\mathrm{A}$), and we check that $L_a$ has indeed reconnected between both times, and that $F_a$ is now located in the leading edge of the QSL footprint. If not, the footpoint $F_a$ is adjusted manually. Next we integrate the field line $L_b(t_i)$ from footpoint $F_c$ at $t=t_i$, and thus we obtain its conjugate footpoint $F_b$ located in the same polarity as $F_a$. We also integrate $L_b(t_f)$ from $F_b$ at $t=t_f$ and we check that $L_b$ has indeed reconnected between both times. If not, $F_a$ has to be adjusted again. Finally, the four field lines $L_u(t_v)$ that correspond to the initial and final states of the ``pairwise reconnection'' can then be (re)integrated at both times $t=t_v$ from both fixed footpoints $F_u$ located in one same polarity (with $u=a;b$ and $v=i;f$). 

It must be recalled that, even though both field lines do exchange one of their conjugate footpoints $F_c$ during the ``pairwise reconnection'', the nature of QSL reconnection still implies that they cannot rigorously exchange their other conjugate footpoint, and that both of these conjugate footpoints slip over some distance before and after this reconnection. This explain the offsets between field line footpoints which are displayed hereafter, even though TOPOTR is an accurate field line integrator. And more generally it warrants caution when analyzing (or sketching) field lines reconnecting in 3D.

   \begin{figure*}
   \centering
   \includegraphics[width=0.85\textwidth,clip]{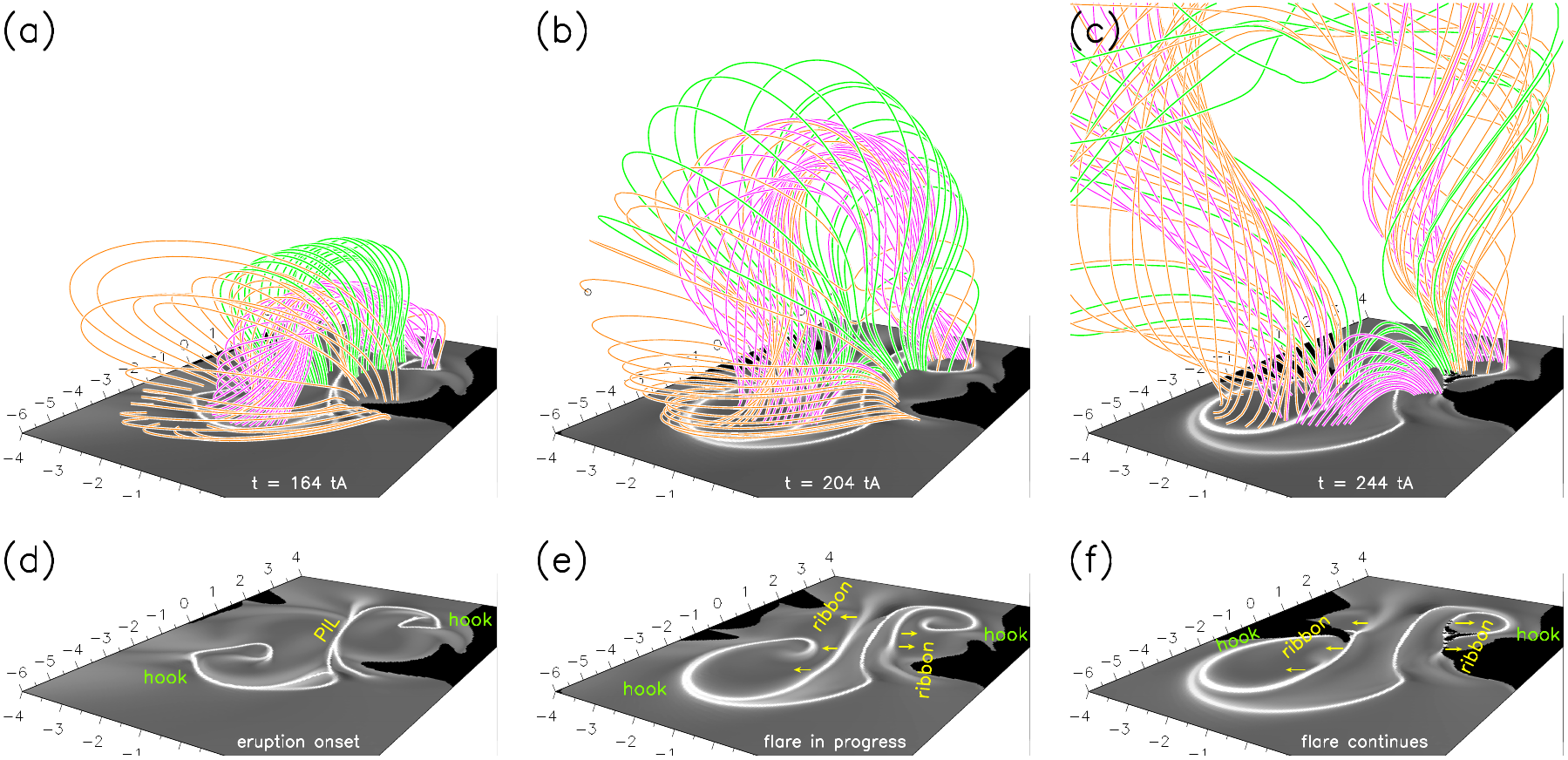}
   \caption{Time-evolution of representative coronal field lines (\textit{top row}) and of 
		the double-J shaped QSL footprints (\textit{bottom row}) during the modeled flare, from 
		the eruption onset (\textit{left}) up to the end of the simulation (\textit{right}). 
		The pink (resp. orange) field lines correspond to the external edge of the flux-rope 
		as it is at the eruption onset (resp. at the end of the simulation). The green field lines 
		correspond to overlaying arcades at the eruption onset. The 
		QSL footprints are represented by the squashing degree $Q$, with a logarithmic 
		greycale corlor coding with grey-white standing for $\log Q = \left< 0; 4 \right>$, and with 
		black areas mapping the footpoints of field lines that leave the full numerical 
		domain. The polarity inversion line (PIL) and the relevant hook and ribbon portions 
		QSL footprints are labeled, and the displacements of the latter are indicated 
		by arrows. 
		}
              \label{figropeevol}%
    \end{figure*}

\section{Drifting of the flux rope}
\label{secrope}

\subsection{High-$Q$ footprints of flux-rope QSLs}
\label{sechooks}

As obtained with earlier topological studies of flux ropes \citep[as initiated by][see Sect.~\ref{secmethqsl}]{Dem96b}, the high-$Q$ region in our simulation displays a double-J pattern at $z=0$ during the modeled eruption. It is plotted in the lower panels of Figure~\ref{figropeevol}. This double-J is composed of two elongated nearly-parallel, straight, ribbon-like components, and each of them is prolonged by one curved hook at its end. 

As the eruption progresses, the values of squashing degrees in the double-J do not evolve significantly. High values $Q>10^5$ are present along most of the length of the QSLs, with a peak at $Q=\mathrm{a~few}~10^7$ in the middle of each curved hook. However their shapes do evolve noticeably. In particular the straight parts of both Js spread away from each other and from the PIL (see Figure~\ref{figropeevol}(e)--(f)). This is a result of the flare magnetic reconnection, which occurs at the hyperbolic flux rube (HFT) in the core of the coronal part of the QSL \citep{Aula12,Jan13,Sav16}. In addition, the hooks change shape. In a first phase their curvature becomes smoother and the area surrounded by the QSL footprints expands (see Figure~\ref{figropeevol}(d)--(e)). Then their expansion continues while the distance of between their end and the edge of the straight parts of the Js decreases, so the distinction between the straight part of the ribbon and its hook becomes difficult (see Figure~\ref{figropeevol}(e)--(f)). 

Following the procedure as described at the end of Sect.~\ref{secmethqsl}, we plot the field lines of the outer edge of the flux rope. Their time-evolution is described and analysed hereafter in Sect.~\ref{secfl}. 

\subsection{QSLs unrelated to the erupting flux rope}
\label{secnoiseqsls}

In addition to the double-J pattern, some high-$Q$ features that are neither related with the flux rope drifting nor to the flare reconnection are also present in the simulation. 

Firstly, the double-J actually has a single S-shape at the eruption onset (see Figure~\ref{figropeevol}(d)). This is due to the formation process of the flux rope through flux cancellation. It generates a bald-patch (BP) along the PIL \citep[as defined in][]{Titov93}, at which continuous reconnection gradually builds the flux rope \citep[as described in][]{vanBalle89}. During the flux-rope formation, the hooks correspond to the ends of a BP separatrix with $Q\rightarrow\infty$ \citep[as in the models by][]{TD99,Titov08} before it transitions into a QSL with finite-$Q$ values \citep[as it occurs in][]{Aula10}. 

\begin{figure}
   \centering
   \includegraphics[width=0.3\textwidth,clip]{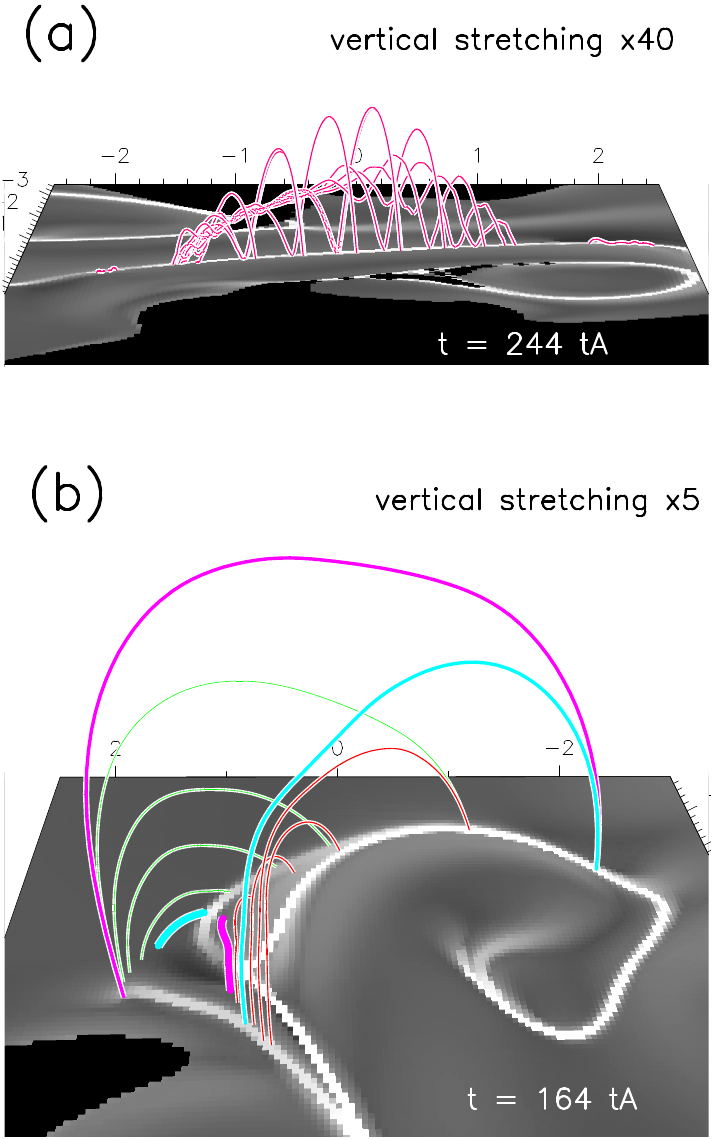}
   \caption{
	{\color{black} Representative field-lines associated with high-$Q$ regions unrelated with 
	the modeled flare and eruptive flux rope. 
	(a): Separatrix field-lines rooted in a bald patch, which outline a 
	small non-eruptive flux rope that results from the splitting of the 
	initial flux-rope in the early phase of the eruption. 
	(b): Field lines which belong to low-$Q$ shear-driven QSLs, which are 
	located on the edge of the main QSLs that surround the erupting flux 
	rope. 
	In both panels a large vertical stretching factor was applied for a better 
	rendering of the geometry of these relatively low-lying and flat structures. 
	} 
		}
              \label{figqslsuseless}%
    \end{figure}

Secondly, a high-$Q$ region remains along the PIL after the flux rope lift-of has begun, between the straight, ribbon-like parts of the QSL-footprints (see Figure~\ref{figropeevol}(e)--(f)). This feature is the same BP as described above. It remains because the eruption leads to low-altitude pinching and reconnection of the BP separatrix. This reconnection leads to ``splitting of the rope in two with one rope successfully being expelled and the other remaining behind'' \citep[as found and coined by][]{Gibss06}.  
{\color{black} In our simulation the X-line that forms underneath the eruptive component originates from a topological bifurcation that produces the HFT (hence the QSL) across which the flare reconnection occurs. This HXT first appears a few grid points only above the line-tied boundary. So the flux rope that is left behind from the eruption and that remains attached to the BP is an extremely low-lying structure, whose radius is only about $1\%$ of its length. This flux rope is represented in Figure~\ref{figqslsuseless}(a), where the vertical axis was stretched by a factor 40 to make the rope visible.} 

The third feature is a couple of lower-$Q$ regions (with $Q\simeq10^{3-4}$) that curve away from both ends of the BP and that gradually elongate towards the edge of the closest hook. These are actually shear-driven QSL that developed long before the eruption onset. The pre-eruptive long-duration dragging of field-lines footpoints from various distances in the asymmetric bipole naturally led to develop connectivity gradients. QSLs thus developed between shorter sheared field-lines rooted near the PIL and longer inclined sheared field-lines that were carried from the outer edge of one polarity towards the PIL. It is interesting to note that these QSLs also involves a (broad) HFT, and that low-altitude reconnection-driven footpoint-exchanges between short and long inclined field-lines also happen there during the simulation. 
{\color{black} Typical field-lines mapping this whole QSL at the onset time of the eruption are plotted in Figure~\ref{figqslsuseless}(b). The overall magnetic-field geometry at this fixed time illustrates how cyan field-lines can later reconnect with one another, as they can slip along the path of green and red thin field-lines and eventually change their connectivities into those of the plotted pink field-lines.}
%

\subsection{Flux rope evolution}
\label{secfl}

Figure~\ref{figropeevol}(a) shows in pink the flux-rope edge at the onset of the eruption, and 
{\color{black} Figure~\ref{figropeevol}(c)}
shows with different orange field lines the flux-rope edge at the end of the simulation. Due to the arched-shape geometry and to the coronal expansion of the flux rope, it is difficult to compare these two sets of field lines with one another. So the field lines originating from the same footpoints and keeping the same colors are plotted on all panels (a)--(c) so as the highlight their changes in connectivity as a function of time. 

The field-line plots reveal that both footpoints of the flux rope have moved away from the PIL between the eruption onset time and the end of the simulation. Given the line-tying of field lines during the eruption, this is not a result of any photospheric motion. Instead Figure~\ref{figropeevol} reveals that the flux rope displacement comes from field line changes in connectivity, due to coronal magnetic reconnection. Indeed, all the late flux-rope edge (orange) field line (from $t$\,=\,244$t_\mathrm{A}$) were actually arcades at pre-eruptive times (at $t$\,=\,164\,$t_\mathrm{A}$). They were bending over the leg of the flux rope. On one side of the PIL they were rooted at the periphery of the QSL hook. On the other side they were rooted relatively close to the PIL. But during the eruption (e.g. at $t = 204 t_\mathrm{A}$), these field lines gradually slipped along the straight portions of the QSL towards the hook, in the slipping and slip-running reconnection regimes. Oppositely, the pre-eruptive flux-rope (pink) field-lines (from $t = 164 t_\mathrm{A}$) that were the closest to the PIL gradually reconnected and formed (flare) loops rooted between both straight parts of QSL footprints. The latter behavior is not specific to our model, since a conversion of long flux-rope field lines into shorter loops near a flux-rope footpoint can also bee seen, for example, in the Fig.~11 from \citet{Amari03b}, although it was not analyzed there. 

On the morphology side, the (pink) flare loops that originated from the pre-eruptive flux rope are nearly undiscernable from other (green) flare loops that resulted from reconnection between large-scale overlaying arcades as in the 2D CSHKP model \citep[and in its 3D extensions by][]{Aula12}. Their main differences are that the former (pink) loops are rooted at the ends of the flaring PIL and tend to be slightly longer, while the latter (green) loops are located above the central part of the PIL and tend to be slightly shorter. Similarly, the (orange) field lines at the edge of flux rope that originated from (very) inclined pre-eruptive arcades do not look so different from the (green) flux-rope envelope field-lines that formed according to the 3D extension of the CSHKP model \citep{Aula12,Dudik14}. Both are rooted in the hooks of the QSLs. But the former (orange) edge field lines tend to be rooted in the part of the hook that is the farthest from the center of the flaring PIL, while the latter (green) envelope field lines are all rooted near the end of the hook that is closer to the center of the bipole. Interestingly, some the pre-reconnection inclined (orange) arcades are carried within one the two large-scale vortices that develop at both flanks of the flux-rope \citep[as studied in][]{Zucca17,Dudik17}. So they can account for so-called imploding loops before they reconnect into the flux rope itself. 

On the connectivity side, our simulation predicts that the photospheric line-tied footpoints of erupting flux-ropes can drift away from the flare site, and that they do so because a series of coronal magnetic reconnections gradually erode the flux rope on its inner side that faces the PIL, in turn producing the endmost flare loops along the PIL, while other reconnections build the rope up on its outer side that faces away from the flaring bipole. This implies that while a CME propagates, and even in the absence of interaction with any open field-lines of the solar wind or with any long-distance active region, the CME footpoints gradually drift away from their source (active) region.  

\subsection{Evolving connections from fixed-footpoints swept by moving-QSLs}
\label{secmulticol}

   \begin{figure}
   \centering
   \includegraphics[width=0.43\textwidth,clip]{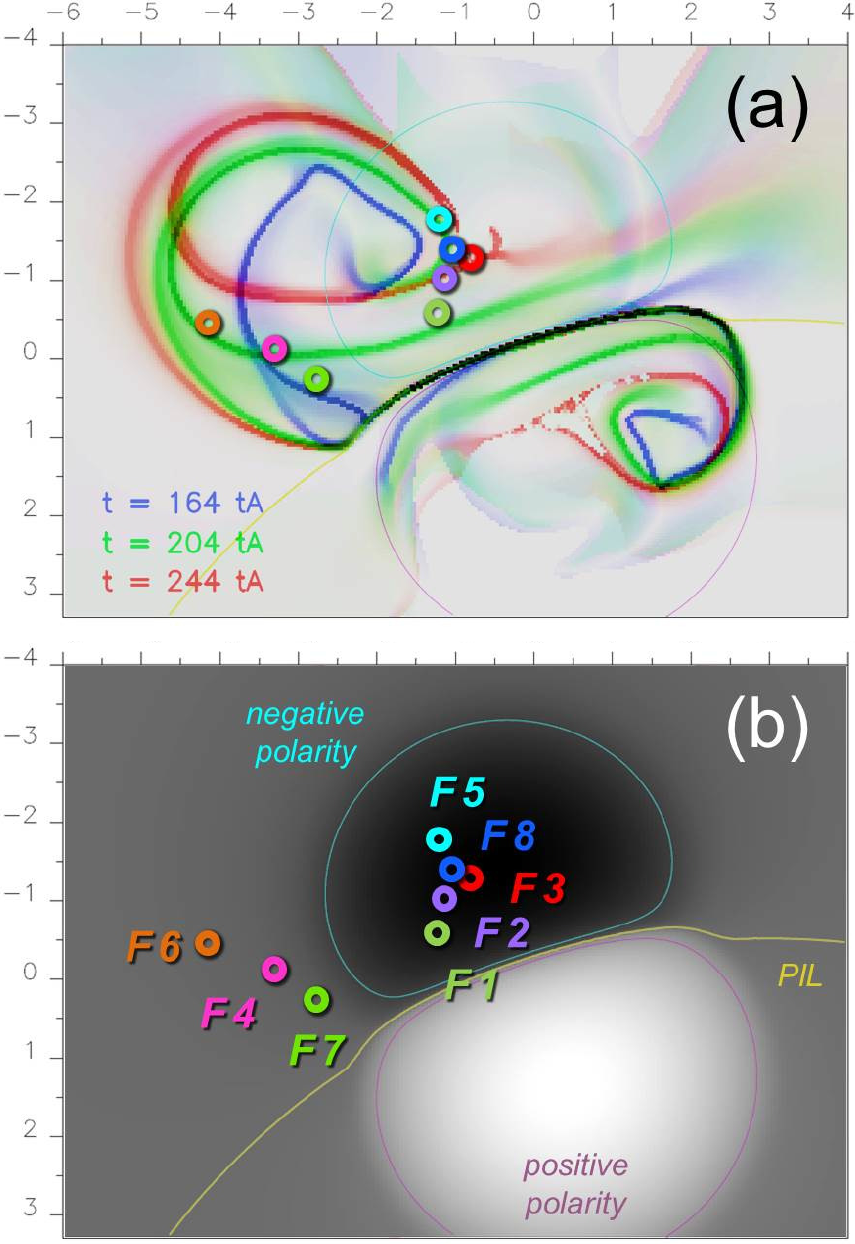}
    \caption{(a): Time-evolution of the double-J shaped QSL footprints (same as in Figure 
	  \ref{figropeevol}) from the eruption onset up to the end of the simulation. (b): Greyscale 
		rendering of the flux distribution in the photosphere, with black-white 
		standing for $B_z = \left<-2.4; 3\right>$. The polarity inversion line 
		(PIL, $B_z = 0$) is shown in yellow in both panels. Similarly, the pink and cyan isoncontours stand for 
		$B_z = 1$ and $-1$, respectively. The colored circles and labels {\it F1--F8} indicate the 
		footpoints in the negative polarity $B_z < 0$ of the field lines whose time-evolutions 
		are plotted in Figures \ref{figreco1}, \ref{figreco2} and \ref{figreco3}. 
		}
              \label{figqslevol}%
    \end{figure}

Here we propose a method to facilitate the understanding of the relations between field-line footpoints, QSL footprints, and magnetic reconnection (as pursued in Sect.~\ref{secreco}), and to establish a link with possible observations of these phenomena (see Sect.~\ref{secobs}). In principle the method that relies on the slip-squashing factors as developed by \citet{Titov09} may be used. But we focus a simpler approach that can also be used readily with EUV observations. We merely overlay the QSL footprints as calculated for different times, and we attribute a different color table for each time so as to distinguish them. The result is displayed in Figure~\ref{figqslevol}. 

This viewing provides a quick and efficient way to observe how the ribbons move away from the PIL and thus turn large-scale arcades into flare loops and flux-rope envelope field-lines, like in standard CSHKP model (as analysed in 3D in Sect.~\ref{secreco1}). For example consider the (purple) footpoint $F2$. For more than half of the simulated eruption (i.e. both at $t = 164 t_\mathrm{A}$ and $t = 204 t_\mathrm{A}$) it remains oustide of the region the PIL that is bounded by the (green) ribbons, as well as of the region encircled by the (blue and green) hook. So the topology of its coronal field line is that of a simple arcade. But at the end of the simulation (at $t = 244 t_\mathrm{A}$) the footpoint is located in the area encircled by the (red) hook. So its field line now belongs to the flux rope, and it ought have reconnected in the second half the modeled eruption (i.e. after $t = 204 t_\mathrm{A}$). 

This method also allows to see where the expansion and displacement of a hook builds up the flux rope in ways that are not predicted by the 2D standard model. One can consider the (cyan) $F5$ foopoint. At the eruption onset is rooted outside the hook so its field line is an arcade. But it must have reconnected (at about $t = 204 t_\mathrm{A}$) when the (green) ribbon reached its position, and it must have turned into the flux rope by the end of the simulation (at $t = 244 t_\mathrm{A}$) since the footpoint is now inside the expanded hook. One can also see the broader hook located in the weaker negative polarity not only drifts much more than the narrower one, but also that one of its sections (around the orange footpoint $F6$) expands and later contracts, which implies the occurrence of multiple reconnection for the same field line (see more in Sect.~\ref{secreco3}). 

Figure~\ref{figqslevol} also highlights which parts of the ribbons gradually take over the hooks, and thus erode the flux rope on its inner side. For example considers the (pink) footpoint $F4$ in Figure~\ref{figqslevol}. At the eruption onset (at $t = 164 t_\mathrm{A}$) this footpoint is encircled by the (blue) hook. So the field line originating from it belongs to the flux rope. But later in the flare (at $t = 244 t_\mathrm{A}$) the ribbon has swept this region, thus resulting in a displacement of the hook away from the footpoint. The field line thus must have reconnected into a flare loop at some intermediate time (right after $t = 204 t_\mathrm{A}$) when the (green) ribbon reached its position (see Sect.~\ref{secreco2}). 

These preliminary inspections also show that the evolution of both hooks can be asymmetric. And they  highlight that the fate of field lines that originate from relatively close footpoints can be very different depending on whether or not they are swept by a QSL footprint, and if it is a hook or a ribbon. We pursue these analyses hereafter. 

\section{New 3D geometries for magnetic reconnection}
\label{secreco}

\subsection{Standard reconnection between arcades}
\label{secreco1}

   \begin{figure*}
   \centering
   \includegraphics[width=0.85\textwidth,clip]{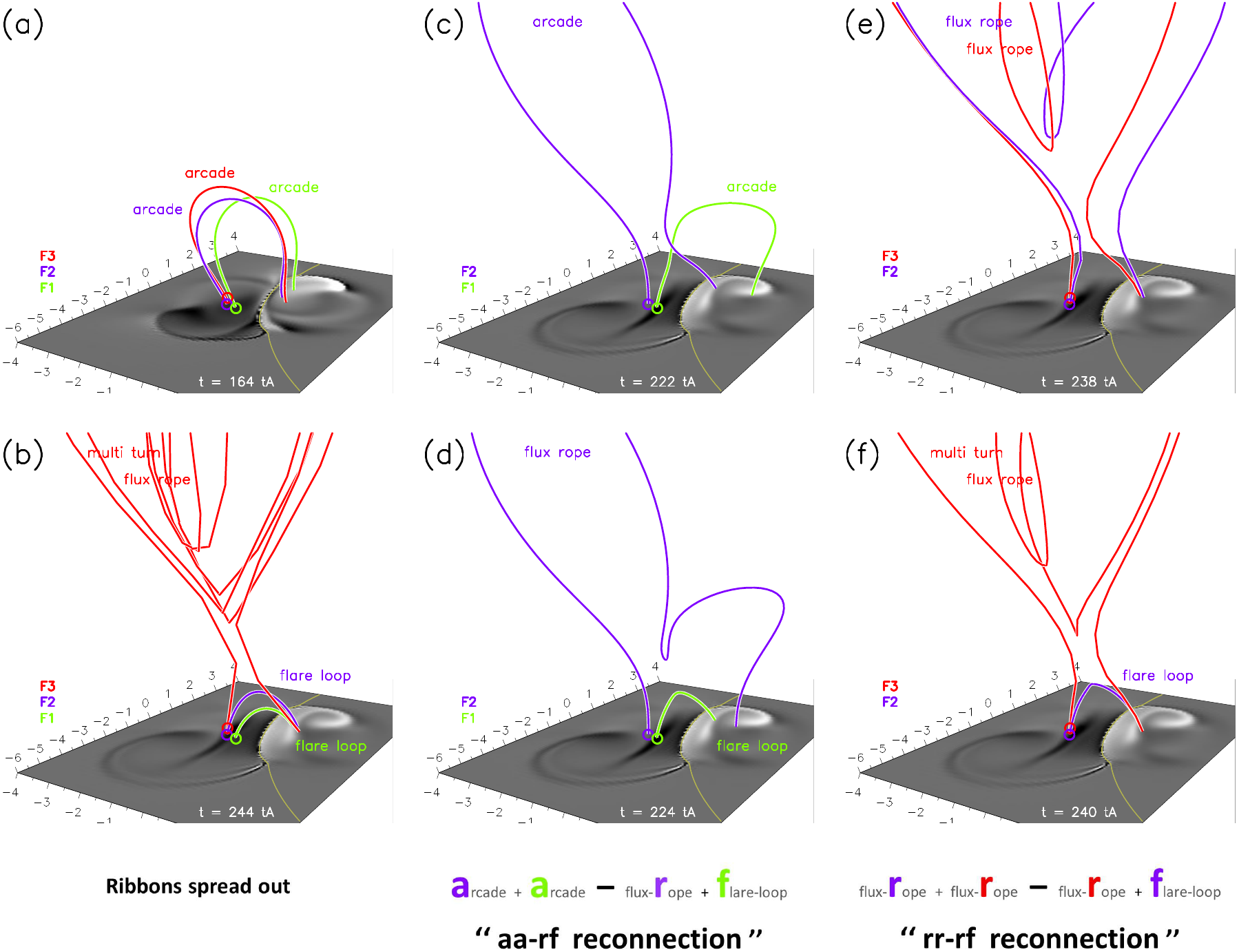}
   \caption{
	  Projection views of the three field lines rooted in the footpoints 
	  {\it F1}, {\it F2} and {\it F3}, plotted at selected times. Each field 
		line is labeled according to its topology at the time that it is plotted, 
		and is colored according to that of its footpoint as plotted in Figure 
		\ref{figqslevol}. The bottom planes show the double-J shaped current ribbons, 
		seen as greyscale rendering of the vertical component of the electric current 
		$\jmath_z$ in the photosphere, with black-white standing for $\jmath_z = \left<-2.4 
		; 3.2\right>$.
		Panels (a) and (b) show the field lines at the eruption onset at $t = 164 t_\mathrm{A}$ 
		and at the end of the simulation at $t = 244 t_\mathrm{A}$. 
		Panels (c)--(d) (resp. (e)--(f)) show the field lines rooted 
		in $F2$ and $F1$ (resp. in $F2$ and $F3$) before and after they endure one 
		``aa-rf'' (resp. one ``rr-rf'') reconnection episode occurring at $t = 223 t_\mathrm{A}$ 
		(resp. at $t = 239 t_\mathrm{A}$). The bottom of the figure indicates which part of 
		the QSL-footpoint (i.e. flare ribbon) evolution is associated with the 
		kinds of reconnections being represented in the right columns. 
		}
              \label{figreco1}%
    \end{figure*}

We analyze the geometry of the reconnection which involves a pair of arcades that initially overlay the flux rope, like it occurs in 2D in the standard CSHKP flare-model. While this was also done in \citet{Aula12} with a similar model, here we also investigate the long-term evolution of the same reconnected field lines. 

Using Figure~\ref{figqslevol} we select a footpoint $F1$ so that its corresponding field line is an arcade at the eruption onset, and is a flare loop at the end of the simulation. These initial and final natures of the field line are ensured by chosing $F1$ to remain outside of the hook at all times, and have it being swept by the straight, ribbon-like portions of the QSL-footprint at some time during the simulation. 

Then using the method as described in Sect.~\ref{secmethreco} we find the footpoint $F2$ of the (purple) field line that reconnects with the (green) one originating from $F1$, and the time at which they reconnect. And follow the time-evolution of these two field lines from the fixed footpoints $F1$ and $F2$ both located on the same side of the PIL. Both arcades are plotted in Figure~\ref{figreco1}(a) at the time of the eruption onset at $t = 164 t_\mathrm{A}$. Their pre-reconnecting shapes at $t = 222 t_\mathrm{A}$, whose expansion results from the eruption of the flux rope underneath, are shown in Figure~\ref{figreco1}(c). Their post-reconnection state at $t = 224 t_\mathrm{A}$ is visible in Figure~\ref{figreco1}(d). These three panels merely show a 3D version of the CSHKP model: the arcade rooted in $F1$ turns into a a flare loop; and the arcade rooted in $F2$ becomes part of the flux rope. Since this reconnection naturally builds up the flux rope envelope, it contributed to an increase in the flux-rope footpoint area, and thus in a growth of the QSL hooks. 

The final state of the system is plotted in Figure~\ref{figreco1}(b). There the expected contraction at $t = 244 t_\mathrm{A}$ of the (green) flare loop originating from $F1$ that results from its slow relaxation during $20 t_\mathrm{A}$ is readily visible. But most surprising is the final state of the (purple) field line rooted in $F2$. This one did not keep expanding within the flux rope as implied in the CSHKP model. Instead it also turned into a flare loop. Further analysis as reported in Figure~\ref{figreco1}(e)--(f) reveals that this flux-rope field-line actually reconnected between $t = 238$ and $240 t_\mathrm{A}$ with another (red) flux-rope field-line rooted in the footpoint $F3$, thus forming a (red) multi-turn flux-rope and turning the (purple) field line into a flare loop. Plotting this new (red) field line at the time of the eruption onset shows that it started as a simple arcade, so it also evolved into a flux rope in a CSHKP-like reconnection (which occurred around $t = 230 t_\mathrm{A}$ and is not shown here). Plotting the (red) field line rooted in $F3$ at the time at which the simulation ends (at $t = 244 t_\mathrm{A}$) reveals that its number of turns have further increased to about 5, so it must have endured multiple reconnections. 

\subsection{New reconnection terminologies}
\label{secrecoterm}

   \begin{figure*}
	 \sidecaption
   \centering
	 \hspace{0.0685\textwidth} 
   \includegraphics[width=0.561\textwidth,clip]{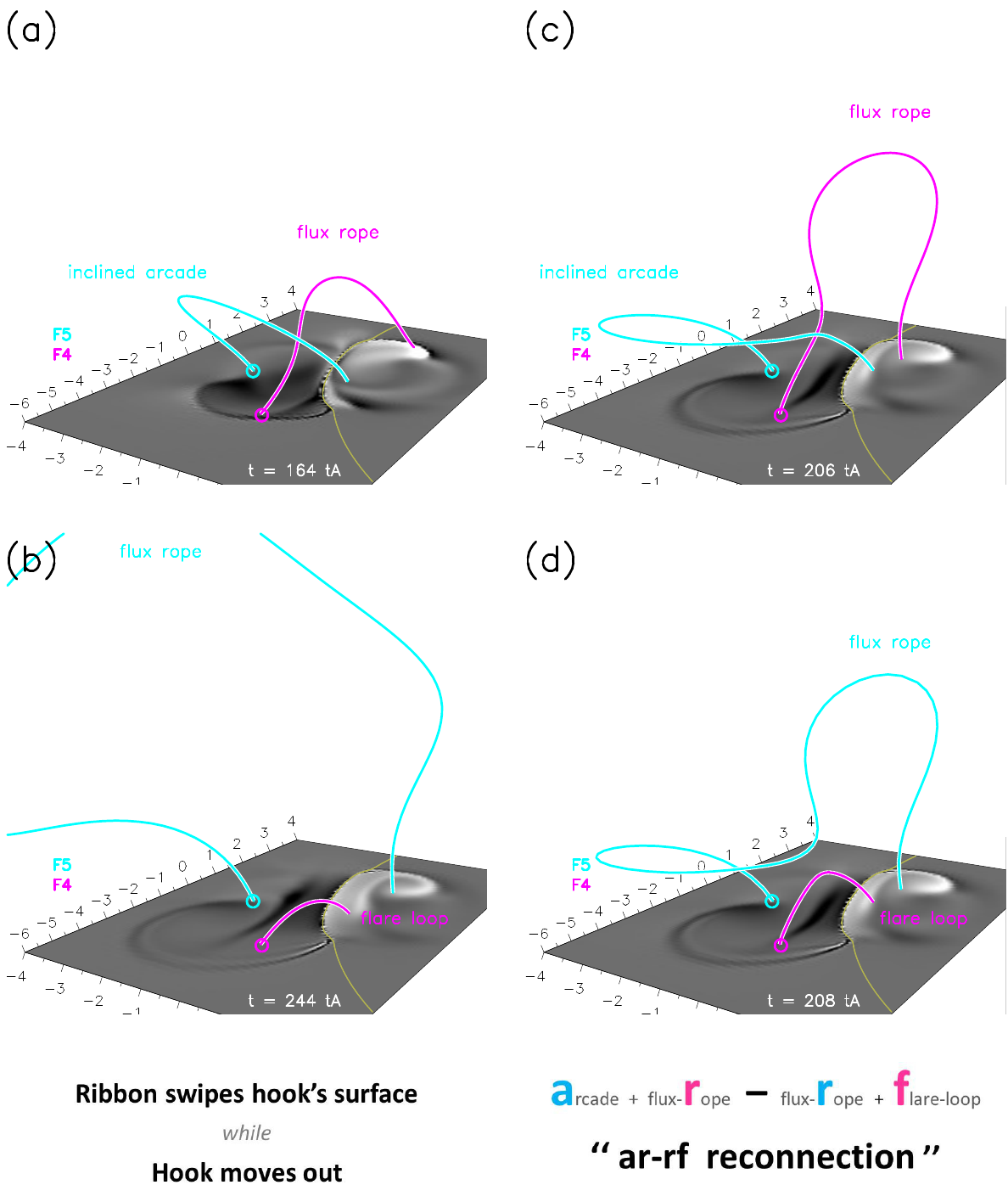}
   \caption{
		%
		Same as Figure \ref{figreco1}, but for the two footpoints {\it F4} and {\it F5}. 
		Panels (a) and (b) show the field lines at the eruption onset at $t = 164 t_\mathrm{A}$ 
		and at the end of the simulation at $t = 244 t_\mathrm{A}$. Panels (c) and (d) show 
		both field lines before and after they endure one ``aa-rf'' reconnection episode 
		occurring at $t = 207 t_\mathrm{A}$. 
	}
              \label{figreco2}%
    \end{figure*}

At this stage we introduce new terminologies so as to describe the geometries that are involved in the various types of reconnection which occur in our eruptive-flare model. 

We define the code-letter {\em a}'' for arcade field lines, the code-letter {\em r}'' for flux-rope field lines, and the code-letter {\em f}'' for the flare loops. We describe the geometries by two pairs of code-letter separated by a hyphen, with the pre (resp. post) reconnection state on the left (resp. right) of the hyphen. When it is required, we respect the order of the code-letters when considering fixed footpoints located in the same polarity, i.e. the first (resp.second) code-letter correspond to the field line that is rooted in the same fixed footpoint as that of the third (resp. fourth) code-letter. 

In this way, the reconnection as shown in Figure~\ref{figreco1}(c)--(d) that involves two arcades that turn into a flux-rope and a flare loop is a {\em aa-rf reconnection}. This is the standard geometry of the standard CSHKP flare model. Using the same rule, the reconnection as shown in Figure~\ref{figreco1}(e)--(f) that involves two flux-rope field lines that reconnect into an another flux-rope field line and a flare loop is a {\em rr-rf reconnection}. 

Hereafter we pursue our analysis on the identification of reconnection geometries that result in the drifting of the flux rope, using this same new terminology. And every type of reconnection is indicated with this terminology at the bottom of the corresponding illustrations, as seen in Figure~\ref{figreco1}. 


\subsection{Reconnection between inclined arcades and rope leg}
\label{secreco2}

   \begin{figure*}
   \centering
   \includegraphics[width=0.85\textwidth,clip]{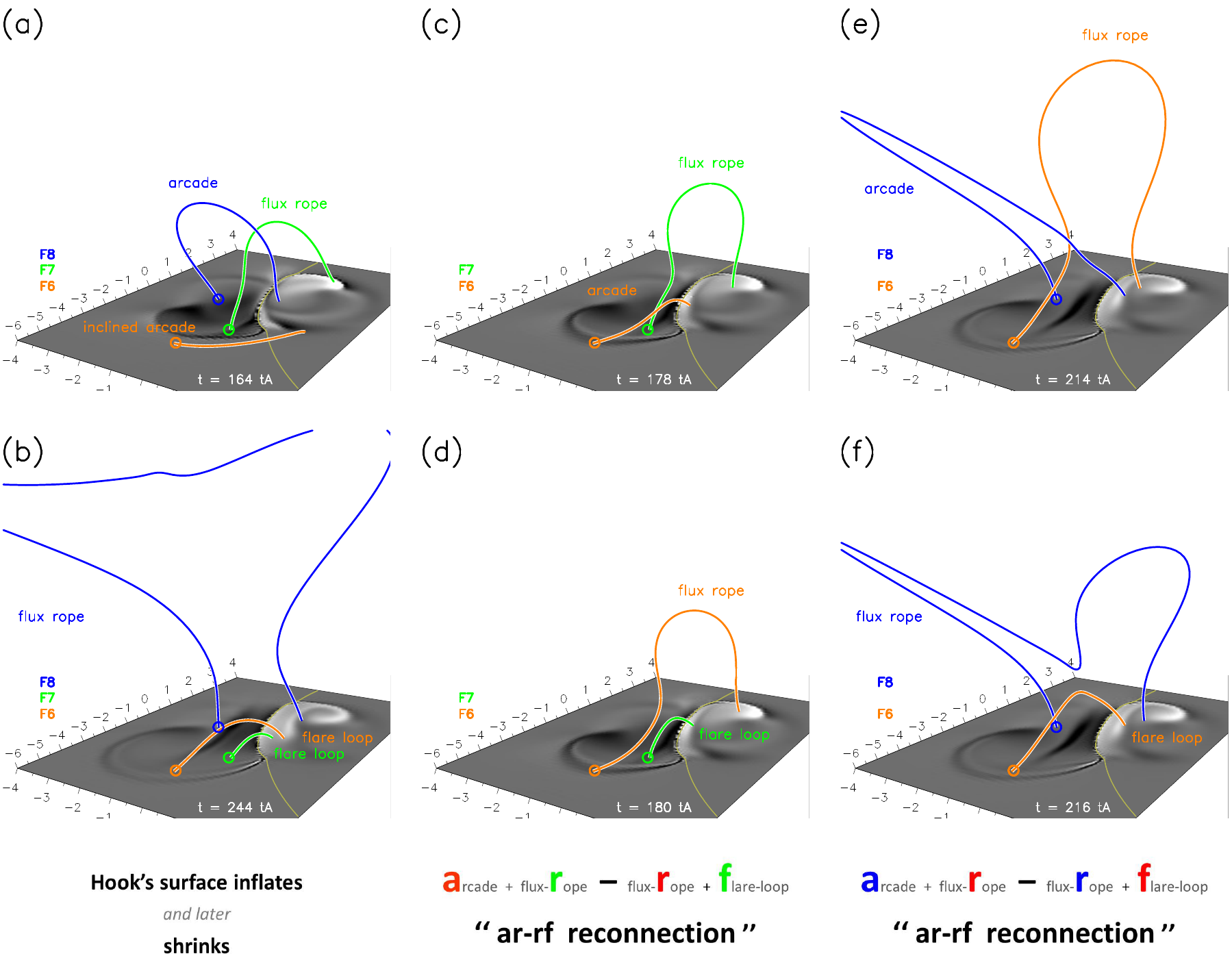}
   \caption{
		%
		Same as Figure \ref{figreco1}, but for the three footpoints {\it F6}, {\it F7} and {\it F8}. 
		Panels (a) and (b) show the field lines at the eruption onset at $t = 164 t_\mathrm{A}$ and 
		at the end of the simulation at $t = 244 t_\mathrm{A}$. Panels (c)--(d) (resp. (e)--(f)) 
		show the field lines rooted in $F6$ and $F7$ (resp. in $F6$ and $F8$) before and after they endure two 
		``aa-rf'' reconnection episodes occurring at $t = 179 t_\mathrm{A}$ (resp. at $t = 215 t_\mathrm{A}$). 
	}
              \label{figreco3}%
    \end{figure*}

We now analyze the type of reconnection that is primarily involved in shifting flux rope footpoints away from the flaring PIL, as previously noted in Section~\ref{secfl}. 

In order to visualise the flux-rope erosion on its inner side that faces the PIL, we select a footpoint $F4$ so that its corresponding field line belongs to the flux-rope at the eruption onset, and later becomes a flare loop. This is just like the reconnecting flare-loop pink field-lines as plotted in Figure~\ref{figropeevol}(c). These initial and final natures of the field line are ensured by choosing $F4$ to be located within the hook at $t = 164 t_\mathrm{A}$, and have it being swept by the straight parts of the QSL-footprint at some time during the simulation. The selected position for $F4$ is shown (in pink) in Figure~\ref{figqslevol}: it is on the inner side of the (blue) hook at the eruption onset, and it is swept by the (green) moving ribbon some short time after to $t = 204 t_\mathrm{A}$. 

We find the footpoint $F5$ of the (cyan) field line that reconnects with the (pink) one originating from $F4$. It is located at the edge of the hook, on its side which is located away from the PIL. Figure~\ref{figreco2}(a) shows that, at the eruption onset, the field line rooted in $F5$ is an inclined arcade that overlies the end part of the flux rope. Figure~\ref{figreco2}(c) (resp. Figure \ref{figreco2}(d)) show the pre (post) reconnecting shapes at $t = 206 t_\mathrm{A}$ ($208 t_\mathrm{A}$) of this pair of field lines. They reveal that the reconnection has transferred the inclined arcade into the flux-rope while it turned the flux-rope field-line into a flare loop. So according to Section~\ref{secrecoterm} this is a {\em ar-rf reconnection}. 

\subsection{Flare loops resulting from multiple reconnections}
\label{secreco3}

As noted in Sect.~\ref{secmulticol}, the leftmost part of the QSL hook as plotted Figure~\ref{figqslevol} shows a non-monotonic drifting. It starts by a (leftward) expansion. And it is followed by a (rightward) contraction. This back and forth motion implies that individual field-lines sequentially move in-and-out of the flux rope through multiple reconnections.  

Here we follow the evolution and analyse the series of reconnections of the field line which is rooted in the footpoint $F6$, chosen to be located in the area which is swept in both directions by the moving hook. Figure~\ref{figreco3}(a) shows that at the onset time of the eruption this (orange) field line is a very low-lying inclined arcade. We follow the same procedure as explained in Sect.~\ref{secmethreco} and as described above, so as to identify both footpoints $F7$ and $F8$ that are involved in the reconnections. Figure~\ref{figreco3}(a) shows that, before these reconnection happen, the two (green and blue) related field-lines belong to the flux rope and to the overlaying arcades, respectively. 

Comparing Figure~\ref{figreco3}(a) and (c), one can notice that the (orange) arcade has endured a preliminary reconnection between $t = 164 t_\mathrm{A}$ and $178 t_\mathrm{A}$, before the occurrence of the first reconnection that we really study here. By linking the current-ribbons as plotted in Figure~\ref{figreco3} with the QSL footprints from Figure~\ref{figropeevol} and \ref{figqslevol}, it can be seen that this preliminary reconnection involved the low-lying shear-driven QSL as described in Sect.~\ref{secnoiseqsls}. However this QSL is not directly related with the evolution of the eruptive flux-rope evolution. Also it is a by-product of the specific formation-process for the pre-eruptive flux-rope which we used so it may not exist in all eruptive flares. And we see here that this reconnection resulted in a relatively small change in field line connectivity for this arcade. Therefore we argue that this preliminary reconnection may not be important in general for the drifting of erupting flux ropes. Still it is worth mentioning it, since it might happen in some real solar events and then explain some complex pattern in hooked flare-ribbons. 

The first (important) reconnection occurs between $t = 178 t_\mathrm{A}$ and $180 t_\mathrm{A}$ (see Figure~\ref{figreco3}(c)--(d)). There the (orange) arcade reconnects with the (green) flux-rope field-line with the exact same {\em ar-rf reconnection} geometry as described above and as reported in Figure~\ref{figreco2}. 

The second reconnection occurs several tens of $t_\mathrm{A}$ later, after the altitude of the apex of the (orange) flux-rope field-line has roughly doubled (see Figure~\ref{figreco3}(c)--(d)). There this line reconnects with the (blue) arcade, which also expanded from the eruption onset time, as a result of the eruption pushing it from below. This second {\em ar-rf reconnection} turns the initially arcade-type and now rope-type (orange field line) into a flare-loop at $t = 216 t_\mathrm{A}$. 

This sequence of reconnections 
eventually produces one (blue) expanding flux-rope field-line and two (orange and green) slowly-relaxing downward-moving flare-loops (see Figure~\ref{figreco3}(b)). 

\subsection{Discussion}
\label{secrecodisc}

The standard CSHKP flare-model in 2D only allows for reconnection between pairs of arcades that initially overlay the erupting flux rope. It produces a new twisted field-line that increases the poloidal flux of the rope, as well as a flare loop underneath the flux rope. Using our terminology as defined in Sect.~\ref{secrecoterm} it can be called {\em aa-rf reconnection}. One example of the 3D version of this reconnection occurring in our simulation is shown in Figure~\ref{figreco1}(c)--(d). Previously published examples can be seen in the Fig.~2 in \citet{Gibss08} and the Fig.~5 \citet{Aula12}. It can also be inferred in many other 3D eruption models \citep[e.g. as reviewed in][]{Aula14,Inoue16,Green18}, and it is usually invoked to interpret the flare loops and ribbons that are observed in the core of erupting active-regions. 

In addition to the standard {\em aa-rf reconnection}, solar eruptions can also involve reconnections between eruptive field-lines and neighboring (or surrounding) loops. They are associated with topological features such as null-points and QSLs that are present in the corresponding potential field. Like for the standard reconnection, these reconnections can happen in 2D \citep[as modeled e.g. in][]{Ant99,Chen00,Ly13,Mas13}. On the observational side, they can account for so-called secondary flare ribbons that develop on the side of (or around) the source region of the flare \citep[e.g.][]{Aula00,Chandra09,Chandra11,Sun13,Sav16}. When modeled in 3D, they can result in the jumping (not the drifting) of one of the flux-rope footpoints to a distant location \citep[e.g.][]{Gibss08,Cohen10,Lugaz11,Jiang13,Vand14}. 

In this work, we have identified two new reconnection geometries and three new reconnection sequences that may occur during eruptive flares, within the erupting bipole independently of any multi-polar topology of the large scale magnetic field. The two new geometries are: 
\begin{itemize}
      \item The {\em rr-rf reconnection}: there two flux-rope field-lines reconnect with each other; this generates a new multi-turn flux-rope field line and a flare loop; the latter is anchored on both sides of the central part of the flaring PIL, between the two straight portions of the QSL footprints (see Figure~\ref{figreco1}(e)--(f)). 
      \item The {\em ar-rf reconnection}: there an inclined arcade rooted within the erupting bipole reconnects with the leg of a flux-rope field line; it generates new flux-rope field line rooted far away from the PIL and a flare loop; the footpoints of the latter are located near the end of the flaring-PIL, one being close to the hooked section of one J-shaped QSL footprint, and the other being near the end of the straight part of the other J-shaped QSL footprint (see e.g. Figure~\ref{figreco2}(c)--(d)). 
\end{itemize}
The three new sequences are: 
\begin{itemize}
      \item An initial standard {\em aa-rf reconnection} followed by a series of {\em rr-rf reconnection}: all together they increase the poloidal flux (i.e. the end-to-end twist) in the outer envelope of the erupting flux-rope; so they increase the area surrounded by the QSL hook (see Figure~\ref{figreco1}). 
      \item A single {\em ar-rf reconnection}: it leads to a field-line footpoint exchange that erodes the flux rope on one side, while it builds it up on the other side; it maintains the total axial flux along the flux rope; it shifts the QSL hook in position and therefore leads to a gradual drifting of the flux rope footpoints (see Figure~\ref{figreco2}). 
      \item A series of the same {\em ar-rf reconnections}: they initially transfer an initially low-lying arcade into the flux rope, and then to turn it into a long flare loop; they shift one part of the QSL hook back-and-forth and eventually also contribute to the drifting of the flux rope footpoints (see Figure~\ref{figreco3}). 
\end{itemize}

All of these reconnections involve the erupting flux-rope and result in the continuous expansion and drifting of its footpoints away from the PIL. They are associated with the QSLs that separate the flux-rope from its surrounding arcades, in particular the QSL hooks, which are purely three-dimensional features. So these reconnections cannot occur in 2D, and are therefore excluded from the standard CSHKP flare model. Nevertheless they eventually produce flare loops, more specifically the ones located at both ends of the flaring PIL. So if the model can be confirmed by observations, then these reconnections will have to be regarded as an important ingredient of eruptive flares. 


\section{Revisiting observations of hooked flare ribbons}
\label{secobs}

\subsection{Two X class flares with hooked ribbons}
\label{secobs1}

To investigate whether the predicted drifting of erupting flux rope footpoints exist in flare observations, we turn to two well-known X-class flares, the X1.4 flare of July 12, 2012 (SOL2012-07-12T16:49) and X1.6 flare of September 10, 2014  \citep[SOL2014-09-10T17:45][]{Dudik16}. We note that these two flares have already successfully served to test the predictions of the 3D extensions to the standard solar flare model \citep{Aula12,Jan13}.

The July 12, 2012 flare was studied by many authors \citep[e.g.,][]{Zhang13,Dudik14,Cheng14,ChengD16,Wang16,Hu16}. Detailed evolution of the ribbons is given by \citet{Dudik14}. A context image of the flare ribbons is shown in Fig. \ref{figobs}(a). This image shows the situation during the flare impulsive phase as observed by the 304\,\AA~filter channel of the Atmospheric Imaging Assembly \citep[AIA][]{Lemen12,Boerner12} onboard the \textit{Solar Dynamics Observatory} \citep[SDO][]{Pesnell12}. The AIA spatial resolution is 1.5$\arcsec$, with a pixel size of 0.6$\arcsec$. The 304\,\AA~passband is dominated by two strong \ion{He}{II} emission lines at 303.8\,\AA~\citep{ODwyer10}. Two bright ribbons, NR and PR, can be readily identified. The PR is shorter and located in the strongest magnetic flux concentration, a pair of leading, positive-polarity sunspots \citep[see Figs. 1 and 2 in][]{Dudik14}. The NR has an extended straight part and a well-formed hook NRH. The hook PRH of the PR extends to the neighboring active regions and is less bright than NRH.

The September 10, 2014  flare is also a well-studied event \citep[e.g.,][]{Cheng15,Graham15,Li15,Tian15,Zhao16,Dudik16,Duan17,Zhang17,Ning17,Lee18}. The evolution of this flare from the precursor to the gradual phase is described in detail in \citet{Dudik16}. The flare ribbons in the impulsive phase of this flare are shown in Fig. \ref{figobs}(b). The ribbon PR is again short, and located in the dominant (leading) positive-polarity sunspot. The conjugate negative-polarity ribbon is more extended, with a straight portion NR and an extended hook NRH. Thus, in both events, the NR and its NRH are located in the weaker-concentrated following negative polarity, similarly as in the simulation, with extended hooks NRH. It is both of these NRH that we focus on from now on.

\subsection{Time-evolution of hooks}
\label{secobs2}

   \begin{figure*}
   \centering
   \includegraphics[width=0.4\textwidth]{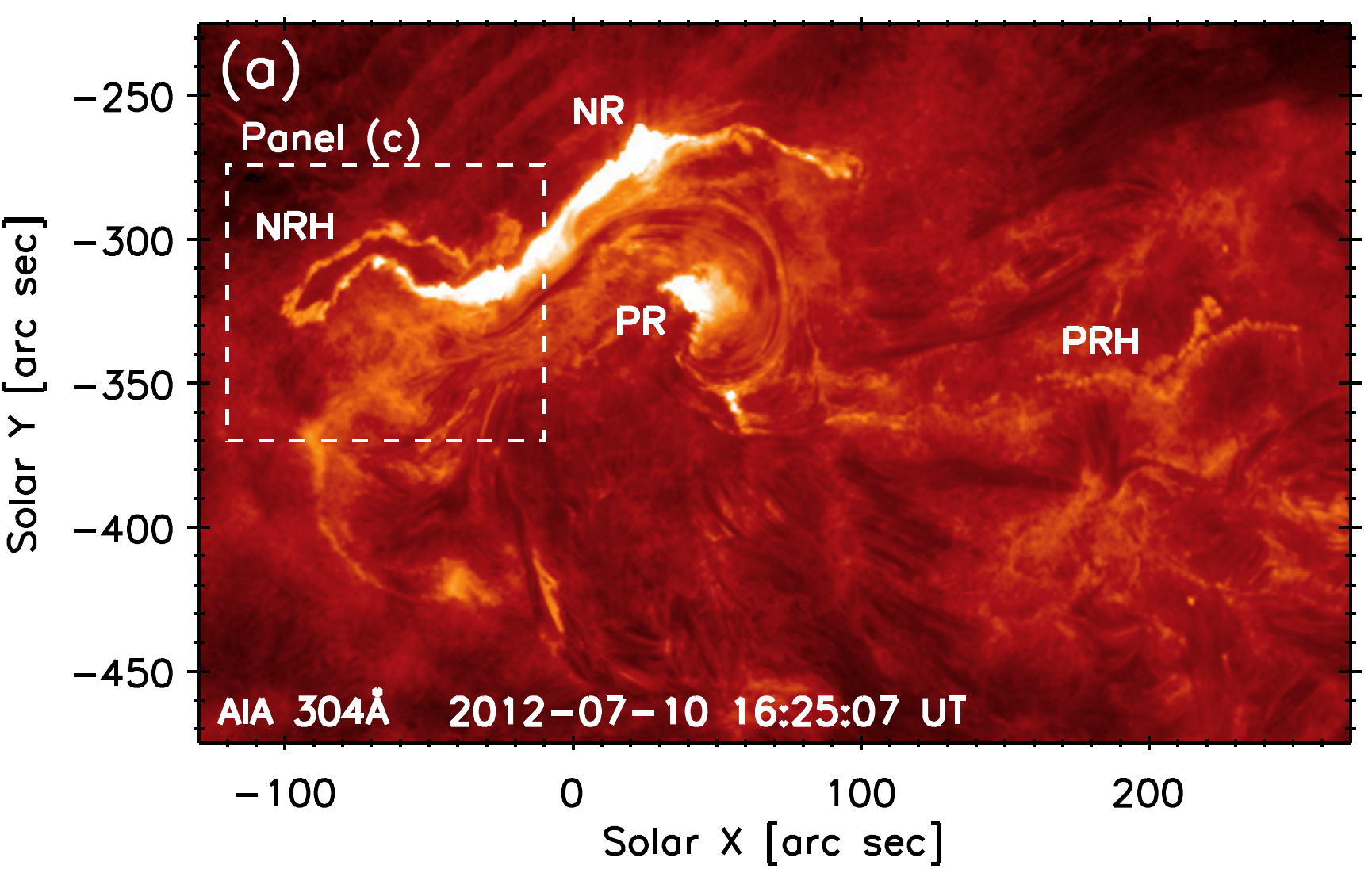}
   \includegraphics[width=0.4\textwidth]{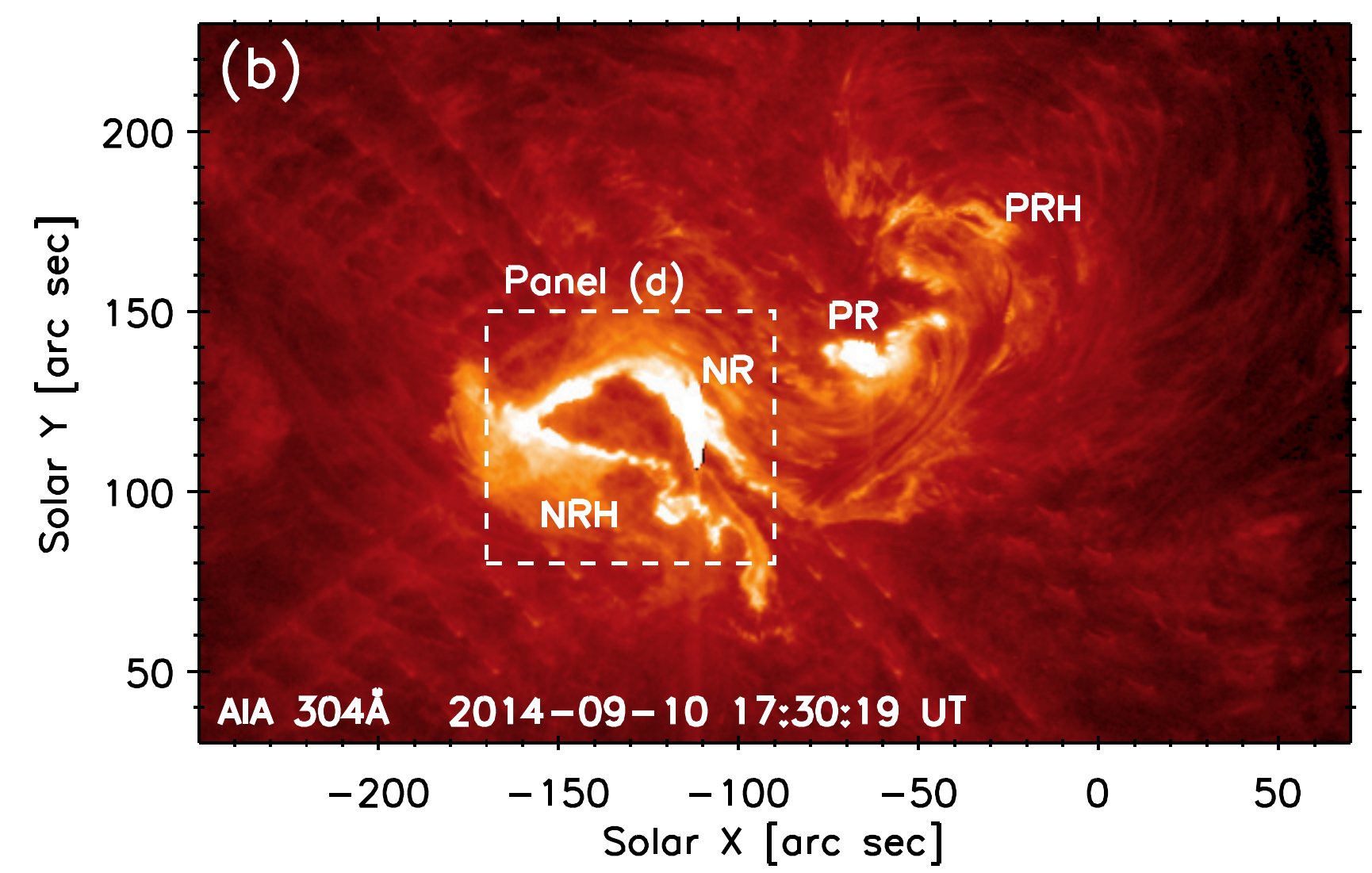}
   \includegraphics[width=0.4\textwidth]{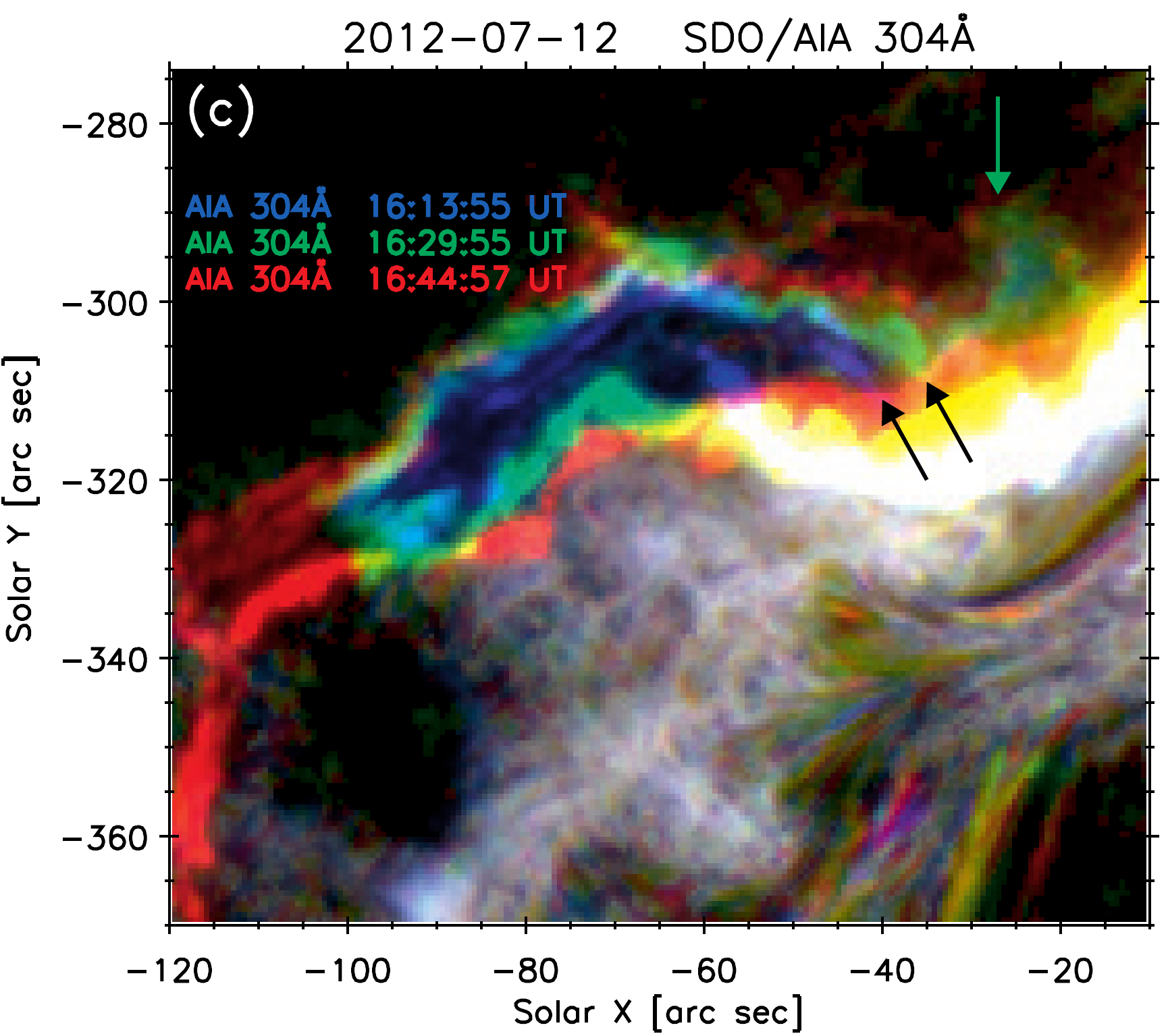}
   \includegraphics[width=0.4\textwidth]{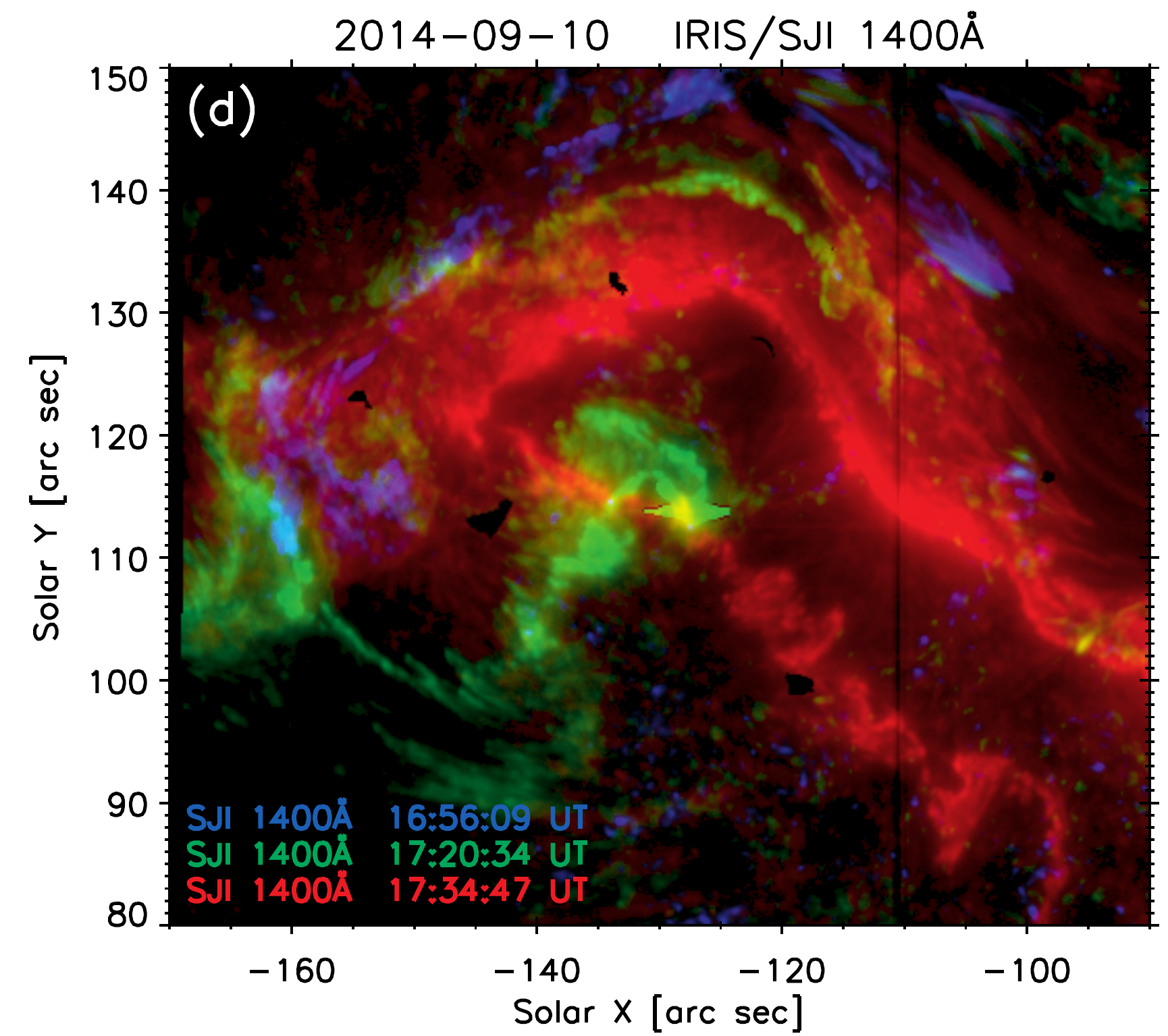}
   \includegraphics[width=0.4\textwidth]{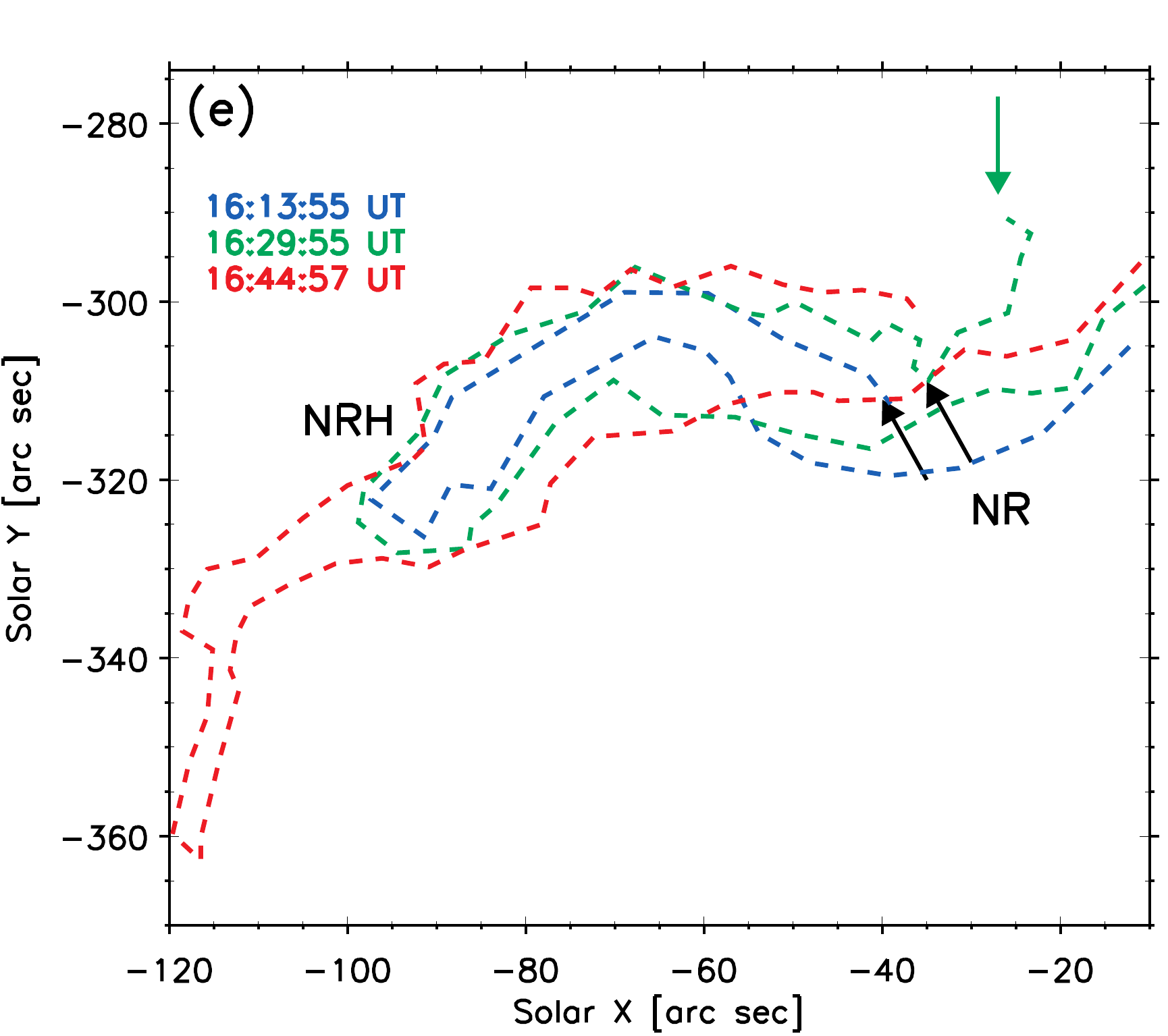}
   \includegraphics[width=0.4\textwidth]{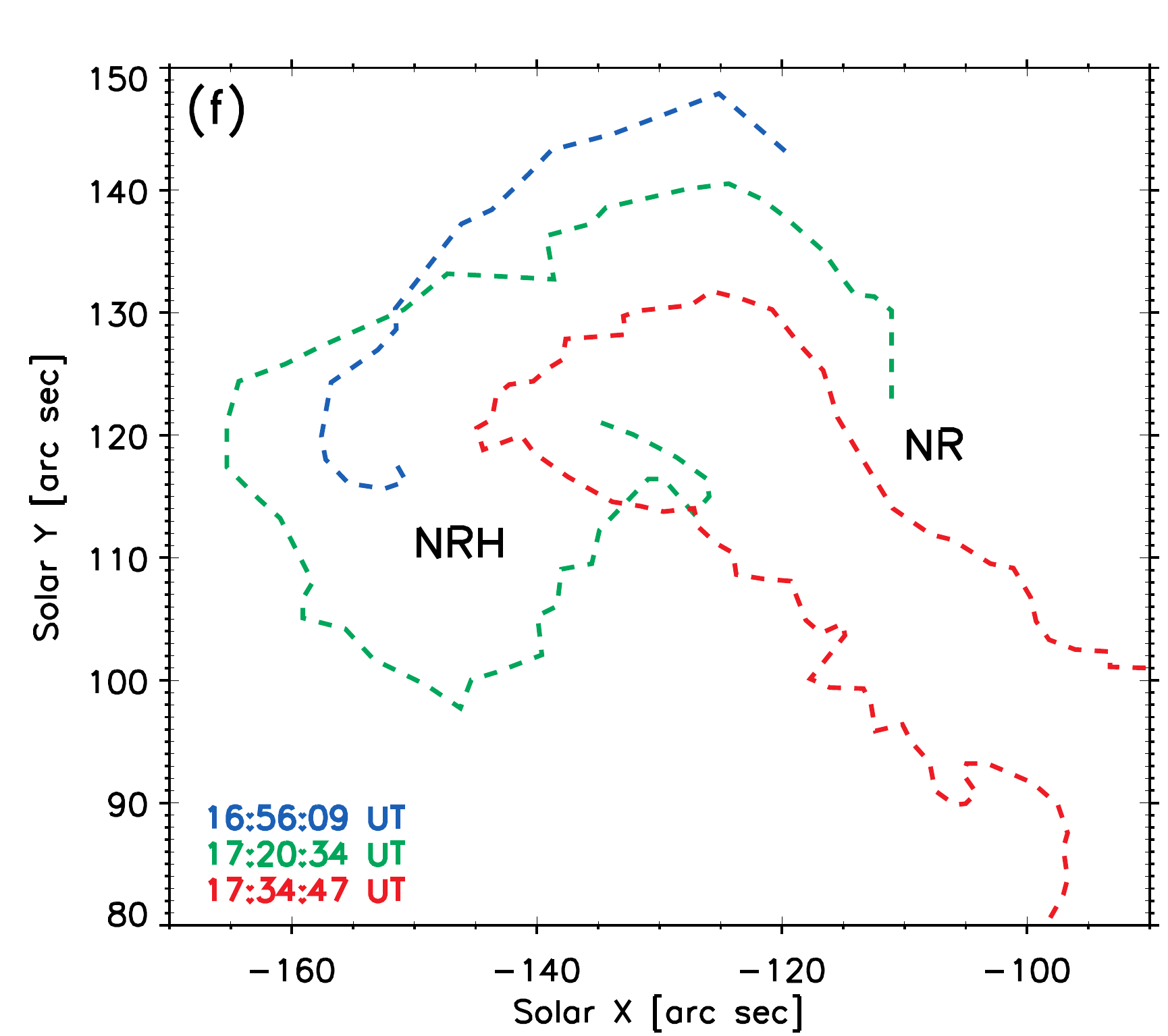}
   \caption{Observations of flare ribbons and their hooks in two X-class flares of July 12, 2012 and September 10, 2014 . Panels (a) and (b) show context images observed in AIA 304\,\AA~during the impulsive phases of each flare. Positive- and negative-polarity ribbons are labeled as PR and NR, respectively, while their hooks are labeled as PRH and NRH. Panels (c) and (d) show zoom-ins on the NRH at three different times, shown by blue, green, and red colors, respectively. Note that in panel (d), the \textit{IRIS}/SJI 1400\,\AA~images are shown rather than AIA 304\,\AA, due to their higher spatial resolution. 
	{\color{black} Panels (e) and (f) show the manually traced-out ribbon locations (dashed colored lines) at times corresponding to panels (c) and (d)}.
	}
              \label{figobs}%
    \end{figure*}
	
The evolution of the NRH in the July 12, 2012 event is shown in the panel (c) Fig. \ref{figobs}. The location of the ribbon NR and its hook NRH is shown at three times, using overlay of blue, green, and red colors. 
{\color{black} In addition, the ribbon shapes at these three times were outlined manually and are shown in panel (e) of Fig. \ref{figobs} by dashed lines of corresponding colors.} 
Blue color displays the situation at the start of the impulsive phase at about 16:14\,UT \citep[see Table 1 of][]{Dudik14}. At this time, an eruption of a hot flux rope rooted in NRH was observed in AIA 131\,\AA~\citep[Fig. 5 in][]{Dudik14}. At this time, the hook NRH is relatively narrow. About sixteen minutes later, at about 16:30\,UT (green color), the NRH has expanded laterally, while its far end have also prolonged by more than a few 10$\arcsec$; 
{\color{black} its tip is now pointed out by the} 
green arrow in Fig. \ref{figobs}(c). We note that the outward expansion of NRH in the north-eastern direction is constrained by the large-scale QSL separating the active region complex from the open field \citep[see Fig. 1 and Sect. 2.1.1 of][]{Dudik14}. At 16:45\,UT (red color), approaching the peak of the flare at 16:49\,UT, the NRH continued its expansion. However, the NR, i.e., the straight portion of the ribbon, moved away from the PIL, sweeping the previous location of the outer edge of NRH (see black arrows in Fig. \ref{figobs}(c)). Thus, the observations of the 2012 July 12 event clearly indicate that this NRH drifts in time. Furthermore, since the tip of the NRH at 16:14\,UT was the location of an erupting S-shape flux rope \citep[see Fig. 5 of][]{Dudik14}, the drifting of the NRH described above indeed indicates the drifting of the footpoints of an erupting flux rope.

Another clear example of a drift of flux rope footpoints can be found during the September 10, 2014 flare. In Fig. \ref{figobs}(d), we show the observations performed by the \textit{Interface Region Imaging Spectrograph} \citep[\textit{IRIS},][]{DePontieu14} in its 1400\,\AA~slit-jaw channel. The emission of the 1400\,\AA~channel is under flare conditions dominated by the \ion{Si}{IV} doublet, originating in transition-region conditions. The reasons for using the \textit{IRIS} observations for this event, instead of \textit{SDO}/AIA 304\,\AA, are twofold. First, during the early (precursor) phase of the flare, the NRH is quite faint, and best visible in \textit{IRIS}/SJI 1400\,\AA. Second, \textit{IRIS} observations allow us to take advantage of their much higher spatial resolution, which is 0.33$\arcsec$.

{\color{black} The \textit{IRIS} 1400\,\AA~observations of the NR and NRH in the 2014 September 10 flare at three different times are shown in Fig. \ref{figobs} (d). The three different times are again shown by the blue, green, and red color. In addition, the ribbon at these three different times is traced out manually in the panel (f) of Fig. \ref{figobs} by dashed lines of respective colors.} 
In the early phase of the flare, the NRH is still growing. At 16:56\,UT (blue color in Fig. \ref{figobs}(d)), the NRH is not yet continous \citep[a feature typical of early phase of flares, see][]{Dudik14,Dudik16}, and quite faint, by about two orders of magnitude compared to the impulsive phase of the flare. We thus had to enhance its intensity for better visibility. At this time, the NRH is only developing; its tip is located at about [$X$,\,$Y$]\,=\,[$-150\arcsec$,\,$112\arcsec$]. However, as the flare progresses towards the impulsive phase, the NRH expands and grows significantly. At 17:20\,UT (green), the NRH expanded eastward and southward, while NR, the straight part of the ribbon, have moved towards SE, i.e., the direction away from the local PIL. We note that in addition to the expansion, the NRH undergoes ``squirming motions'', first noted by \citet{Dudik16}, see Sect. 2.3 and Table 1 therein. These squirming motions accompany the slipping reconnection occurring along the evolving NRH. They indicate that the ribbon passes through a given spatial position more than once. In light of our results, such squirming motions can be explained by multiple reconnections being endured by a field line starting from the same photospheric footpoint.

However, after the onset of fast eruption (at about 17:27\,UT), the NRH starts to shrink and now moves in the westward direction, rather than in the southeastern one. Meanwhile, the straight part NR continues its southeastern motion away from the PIL. At 17:35\,UT (red color in Fig. \ref{figobs}(d)), the NRH is much narrower, while being very bright (more than $\approx$10$^{4}$ DN\,s$^{-1}$\,px$^{-1}$), with trailing bright footpoints of newly-formed flare loops. Thus, the overall evolution of the ribbon, which in this case is not constrained by a large-scale QSL as in the case of the July 12, 2012  event, is in good agreement with the model-predicted evolution. In particular, the area around [$X$,\,$Y$]\,=\,[$-150\arcsec$,\,$112\arcsec$] is first outside of the NRH (at 16:56\,UT, blue color), then becomes a part of the NRH (it is inside NRH at 17:20\,UT; green color), and is outside of the NRH again at 17:34\,UT (red color). A field line rooted in this location therefore had to undergo at least two reconnections, similar as the orange field line starting from the footpoint F6 in Figs. \ref{figqslevol} and \ref{figreco3}.

\subsection{Discussion}
\label{secobs3}

Based on the results of Sect. \ref{secobs2}, we conclude that these two events provide evidence in support of the drifting of flux rope footpoints. However, we note that the evolution of the ribbons alone do not elucidate the individual types of reconnection presented in Sect. \ref{secreco}. This would require analysis of the coronal loops anchored near ribbons (if visible), and their evolution in conjuction with the evolution of the ribbons. This is outside of the scope of this paper, and is left for future work. However, previous observational reports indicating reconnection between open and closed loops \citep[for example,][]{Savage12,Zhu16} already permit for optimism in this regard.

\section{Conclusions}
\label{secccl}

In this paper we jointly addressed the three following questions about solar eruptions: Are the footpoints at the Sun's surface of (I)CME flux ropes located at the same locations as those of their progenitor current-carrying magnetic fields? Do flare loops that develop at the endings of flaring PILs form in the same way as described in the CSHKP standard model?

To address these questions, we performed new analyses of an idealized MHD simulation of solar eruptions, in which the eruption is triggered by torus instability of a flux rope previously formed by flux-cancellation. We built upon earlier extensions of the standard flare model in three dimensions, in particular the fact that the foopoints of eruptive flux-ropes should be located within the hooks of flare ribbons that develop at the footprints of current-carrying QSLs. And we analyse the time-evolution of ribbons in two X-class eruptive-flares observed by \textit{SDO} and \textit{IRIS}.

Our results concerning the development of the standard flare model in 3D can be summarized as follows:
			In spite of line-tying, the footpoints of the modeled flux-rope drift away from the flaring PIL during the  eruption. This is due to a series of coronal reconnections that erode the side of the flux rope facing the PIL, and build it up on its other side facing away from the PIL.
			Several reconnection geometries and sequences being involved have been identified. They are intrinsically three-dimensional so they cannot be described by the CSHKP model. New terminologies were proposed to describe these geometries. In particular the {\em ar-rf reconnection} involves an [{\em a}]rcade and the flux-[{\em r}]ope field line and produces a new flux-[{\em r}]ope field line and a [{\em f}]lare loop.
			These reconnections all together generate a displacement and a deformation of the double-J shaped QSL footprints, in particular of their hooks within which the flux-rope is rooted.
			The flare loops that form at the endings of the flaring PIL do not form by pairwise reconnections of coronal arcades as others do. Instead they are formed by {\em ar-rf reconnection} and are thus are directly related with the flux-rope drifting.
			The evolutions of the hooks of flare ribbons in the two well-known X-class flares of July 12, 2012 and September 10, 2014 are qualitatively consistent with those of the drifting QSL-footprints in our model.

These flare-related behaviors can be linked to the identification of the footpoints of CME and ICME flux-ropes as follows:
			The drifting of the eruptive flux-rope footpoints implies that (I)CME flux-ropes should not map down to the positions of the pre-eruptive flux rope. So magnetic field extrapolations and direct-imagery of the pre-eruptive current-carrying flux tube are not sufficient to identify the locations of (I)CME footpoints.
			Nevertheless the relative positioning of flare ribbons with respect to those of line-tied footpoints may be used to identify the time-evolving topology of field-lines, in particular to follow the drifting footpoints of (I)CMEs flux-ropes within the areas surrounded by the evolving hooks of flare ribbons. 

The flare-related behaviors which we modeled may constitute new extensions to the standard flare model in three dimensions. We already provided some qualitative observational support from the spatio-temporal properties of hooked flare ribbons in two observed events. But more observations are required so as to test further all the model predictions, in particular regarding CME footpoint locations and flare-loop formation at the ends of flaring PILs. 


\begin{acknowledgements}
The authors thank Miho Janvier, Juraj L\"{o}rin\v{c}\'{i}k and Alena Zemanov\'{a}, and for discussions. 
G.A. thanks the Programme National Soleil Terre of the CNRS/INSU for financial support, as well as the Astronomical Institute of the Czech Academy of Sciences in Ond\v{r}ejov for financial support and warm welcome during his visits. 
J.D. acknowledges financial support from the project 17-16447S of the Grant Agency of Czech Republic as well as insitutional support RVO: 67985815 from the Czech Academy of Sciences. 
The simulation used in this work was executed on the HPC center MesoPSL which is financed by the R\'egion Ile-de-France and the project Equip@Meso of the PIA supervised by the ANR. 
AIA data are provided courtesy of NASA/\textit{SDO} and the AIA science team. 
\textit{IRIS} is a NASA small explorer mission developed and operated by LMSAL with mission operations executed at NASA Ames Research Center and major contributions to downlink communications funded by the Norwegian Space Center (NSC, Norway) through an ESA PRODEX contract.
\end{acknowledgements}

\bibliographystyle{aa} 
\bibliography{aulanierREF}  
\IfFileExists{\jobname.bbl}{} {\typeout{}
\typeout{***************************************************************}
\typeout{***************************************************************}
\typeout{** Please run "bibtex \jobname" to obtain the bibliography} 
\typeout{** and re-run "latex \jobname" twice to fix references} 
\typeout{***************************************************************}
\typeout{***************************************************************}
\typeout{}}

%

\end{document}